\documentclass[12pt]{article}
\usepackage{amsmath,amsthm,amssymb,amscd,epsfig,}

\newtheorem{theorem}{Theorem}[section]

\newtheorem{definition}[theorem]{Definition}

\def\e{\mathrm{e}}
\newcommand{\bbbone}{\mathchoice {\rm 1\mskip-4mu l} {\rm 1\mskip-4mu l}
{\rm 1\mskip-4.5mu l} {\rm 1\mskip-5mu l}}

\def\N{{\mathbb N}}
\def\Z{{\mathbb Z}}
\def\R{{\mathbb R}}
\def\C{{\mathbb C}}

\def\d{{\mathrm d}}
\def\e{{\mathrm e}}

\def\H{{\mathcal H}}

\def\I{{\mathrm i}}
\def\K{{\mathcal K}}
\def\Kc{{\mathcal K}^\C}
\def\L2r{L^2_{\mathrm r}}
\def\Ltr{\L2r}

\def\S{{\mathcal S}}

\def\Wf{W_{\mathrm{F}}}

\def\eff{{\cal F}}

\begin{document}
\date{\today}

\author{
Stephan De Bi\`evre\\ UFR de Math\'ematiques et UMR P. Painlev\'e\\
Universit\'e des Sciences et Technologies de Lille\\ 59655 Villeneuve d'Ascq
Cedex France\\e-mail: Stephan.De-Bievre@math.univ-lille1.fr }

\title{Where's that quantum?}

\maketitle

\abstract{
The nature and properties of the vacuum as well as the meaning and localization properties of one or many particle states have attracted a fair amount of attention and stirred up sometimes heated debate in relativistic quantum field theory over the years. I will review some of the literature on the subject and  will then show that these issues arise just as well in non-relativistic theories of extended systems, such as free bose fields. I will argue they should as such not have given rise either to surprise or to controversy. They are in fact the result of the misinterpretation of the vacuum as ``empty space'' and of a too stringent interpretation of field quanta as point particles. I will in particular present a generalization of an apparently little known theorem of Knight on the non-localizability of field quanta, Licht's characterization of localized excitations of the vacuum, and explain  how the physical consequences of the Reeh-Schlieder theorem on the cyclicity and separability of the vacuum for local observables are   already perfectly familiar from non-relativistic systems of coupled oscillators. 
}

\maketitle

\section{Introduction}
Quantum field theory is the study of quantum systems with an infinite number
of degrees of freedom. The simplest quantum field theories are the free bose
fields, which are essentially assemblies of an infinite number of coupled
oscillators. Examples include the quantized electromagnetic field, lattice
vibrations in solid state physics,
and the Klein-Gordon field, some of which are relativistic while others are not. Particles show up in quantum field theory as ``field quanta.''
As we will see, these do not quite have all the properties of the usual particles of
Newtonian physics or of Schr\"odinger quantum mechanics. I will in particular show (Section \ref{s:whethaqua}) that the quanta of free bose fields, relativistic or not, can not be 
perfectly localized in a bounded subset of space.  This, in my opinion, shows conclusively that the difficulties encountered when attempting to define a position observable for the field quanta of relativistic fields, that continue to be the
source of regular debate in the literature \cite{he1} \cite{he2} \cite{he3}
\cite{bibi} \cite{by} \cite{flbu} \cite{hal1} \cite{clha2} \cite{wa1} \cite{wa2}, do not find their origin in any form of causality violation, as seems to be generally thought. Instead, they result from   an understandable but ill-fated attempt to force too stringent a particle interpretation on the states of the quantum field containing a finite number of quanta. 

States of free bose fields that are perfectly localized in bounded sets exist, but necessarily contain an infinite number of quanta. They can be classified quite easily, following the ideas of Licht \cite{li1}. This is done in Sections \ref{s:thingslocal} and \ref{s:chalocsta}. It turns out that those states do not form a vector subspace of the quantum Hilbert space. This has on occasion been presented as surprising or paradoxical within the context of relativistic quantum field theory, but it isn't: these properties of localized states are familiar already from finite systems of coupled harmonic oscillators.

The previous issues are intimately related to certain properties of the vacuum of free bose fields, that also have stirred up animated debate in the literature. I will in particular explain the  physical implications of the Reeh-Schlieder theorem for free bose fields and their link with the localization issue. The Reeh-Schlieder theorem has been proven in the context of relativistic field theory, but holds equally well for (finite or infinite) non-relativistic systems of coupled oscillators. By studying it in that context, one easily convinces oneself that its implications for the quantum theory of measurement, for example, have nothing particularly surprising or paradoxical, and do not lead to causality violation, but are the result of the usual ``weirdness'' of quantum mechanics, since they are intimately linked to the observation that the vacuum is an entangled state. This thesis is developed in Section \ref{s:reesch}. 

The paper is organized as follows. After defining the class of free bose fields under consideration (Section \ref{s:freebosfie}), I will briefly discuss the question of the ``localizability'' of states with a finite number of quanta in Section \ref{s:whethaqua}. I will in particular recall from  \cite{db06} a generalization to free bose fields of a little known result of Knight about the Klein-Gordon field \cite{kn}, which shows they can not be perfectly localized excitations of the vacuum.   I will then discuss the various, sometimes contradictory viewpoints on the question of particle localization prevalent in quantum field theory textbooks in
Section \ref{s:newwigc} and compare them to the one presented here. I will argue in detail that the latter could help to clarify the situation considerably.

Section \ref{s:thingslocal} contains the precise definition of ``localized excitation of the vacuum'' that I am using and some technical material needed for the presentation of Licht's theorem on the characterization of local states in Section \ref{s:chalocsta}. Section \ref{s:reesch} explains the  physical implications of the Reeh-Schlieder theorem for free bose fields and their link with the localization issue.

Some of the proofs missing here, as well as a detailed critique of the Newton-Wigner position operator from the present viewpoint on localization can be found in \cite{db06} which, together with the present work, is itself a short version of Chapter 6 of \cite{db06bis}. I refer to these references for more details.


\section{Free bose fields}\label{s:freebosfie}
The simple physical systems under study here obey an equation of the form
\begin{equation}\label{eq:freeosc}
\ddot q +\Omega^2 q = 0,
\end{equation}
where $\Omega$ is a self-adjoint, positive operator on a dense
domain $\mathcal D (\Omega)$ in a real Hilbert space $\mathcal K$
and having a trivial kernel. I will always suppose $\K$ is of the form $\K = \Ltr(K,
\d \mu)$, where $K$ is a topological space and $\mu$ a Borel
measure on $K$. Here the subscript ``r'' indicates that we are
dealing with the real Hilbert space of real-valued functions. In
fact, all examples of interest I know of are of this type. Those include:

(i) Finite dimensional systems of coupled oscillators, where $\K=\R^n$ and
$\Omega$ is a positive definite matrix;

(ii) Lattices or chains of coupled oscillators, where $\K=\ell^2(\Z^d, \R)$
and $\Omega^2$ is usually a bounded finite difference operator with a possibly unbounded inverse;

(iii) The wave and Klein-Gordon equations, where $\K= L^2(K, \R)$, $K\subset\R^d$
and $\Omega^2=-\Delta+m^2$ with suitable boundary conditions. More precisely, the Klein-Gordon equation is the equation
$$
\partial^2_tq(t,x) = -(-\Delta + m^2)q(t,x).
$$

Equation (\ref{eq:freeosc}) can be seen as a Hamiltonian system with phase space
$$
\mathcal H = \K_{1/2} \oplus \K_{-1/2},\quad{\mathrm{where}}\quad 
\K_{\pm 1/2} = [\mathcal D(\Omega^{\pm 1/2})].
$$
Here the notation $[\ ]$ means that we completed $\mathcal D$ in the topology
induced by $\parallel\Omega^{\pm 1/2} q\parallel$ where $\parallel \cdot
\parallel$ is the Hilbert space norm of $\K$. On $\H$, the Hamiltonian $(X=(q,p)\in \mathcal H)$
\begin{equation}\label{eq:freehamiltonian}
H(X)= \frac{1}{2} p\cdot p + \frac{1}{2} q\cdot \Omega^2 q,
\end{equation}
defines a Hamiltonian flow
with respect to the symplectic structure
$$
s(X,X') = q\cdot p' - q'\cdot p.
$$
The Hamiltonian equations of motion $\dot q= p,\ \dot p =-\Omega^2 q$ are equivalent to
(\ref{eq:freeosc}). Note that I use $\cdot$ for the inner product on $\K$. 

The quantum mechanical description of these systems can be summarized as follows. Given a harmonic system
determined by $\K$ and $\Omega$, one chooses as the quantum Hilbert space of such a system the symmetric Fock
space $\eff^+(\K^\C)$, and as quantum Hamiltonian the second quantization of $\Omega$: $H=\d
\Gamma(\Omega)$. Note that this is a positive operator and that
the Fock vacuum is its ground state, with eigenvalue $0$.
In terms of the standard creation and annihilation operators on this Fock space,
the quantized fields and their conjugates are then {\em defined}
by ($\eta\in\K_{-1/2}^\C$):
\begin{equation}\label{eq:fieldq3}
\eta\cdot Q := \frac{1}{\sqrt2} (a(\Omega^{-1/2}\overline\eta) +
a^\dagger(\Omega^{-1/2} \eta)),
\end{equation}
and, similarly ($\eta\in\K_{1/2}^\C$),
\begin{equation}\label{eq:conjugatefieldq3}
\eta\cdot P := \frac{\I}{\sqrt2}
(a^\dagger(\Omega^{1/2}\eta)-a(\Omega^{1/2}\overline\eta) ).
\end{equation}
For later purposes, I define, for each $\xi\in\K^\C$, the Weyl operator
\begin{equation}\label{eq:weylop}
\Wf(\xi)=\exp(a^\dagger(\xi)-a(\xi)).
\end{equation}
I will also need, for each $X=(q,p)\in\H$
\begin{equation}\label{eq:zomega}
z_\Omega(X)=\frac1{\sqrt2}(\Omega^{1/2}q+\I\Omega^{-1/2}p)\in\K^\C.
\end{equation}
It follows easily that
$$
\Wf(z_\Omega(X))=\exp-\I(q\cdot P-p\cdot Q).
$$
Loosely speaking, the observables of the theory are ``all functions of $Q$ and $P$.'' For mathematical precision, one often uses various algebras (called CCR-algebras) generated by the Weyl operators $\Wf(z_\Omega(X))$, $X\in\H$, as we will see in some more detail below. 

Things are particularly simple when the set $K$ is discrete, as in examples (i) and (ii) above. One can then define the displacements $Q_j$ and momenta $P_j$ of the individual oscillators, with $j=1, \dots, n$ in the first case and $j\in \Z^d$ in the second case. These examples are a helpful guide to the intuition, as we will see below.

Apart from the vacuum $|0\rangle$, which is the ground state of the system, excited states of the form
$$
a^\dagger(\eta_1)\dots a^\dagger(\eta_k)|0\rangle,
$$
play a crucial role and are referred to as states with $k$ ``quanta.'' 
The quanta of the Klein-Gordon field, for example are thought of as spinless particles of mass $m$.  In the case of an oscillator chain or lattice, they are referred to as ``phonons.''  To the extent that these quanta are thought of as ``particles,'' the question of their whereabouts is a perfectly natural one. It is to its discussion I turn next.


\section{So, where's that quantum?}\label{s:whethaqua}
Let me start with an informal discussion of the issue under consideration. Among the interesting observables of the oscillator systems we are
studying are certainly the ``local'' ones. I will give a precise
definition in Section \ref{s:thingslocal}, but thinking for example of a finite or infinite 
oscillator chain, ``the displacement $q_7$ or the momentum $p_7$
of the seventh oscillator'' is certainly a ``local'' observable.
In the same way, if dealing with a wave equation, ``the value
$q(x)$ of the field at $x$'' is a local observable. 
Generally, ``local observables'' are functions of the
fields and conjugate fields in a bounded region of space. In the
case of the oscillator chain or lattice, space is $K=\Z^d$ $(d\geq 1)$, and
for a finite chain, space is simply the index set $K=\{1,\dots
n\}$.

I find this last example personally most instructive. It forces
one into an unusual point of view on a system of $n$ coupled
oscillators that is well suited for making the transition to the
infinite dimensional case. Think therefore of a system of $n$
oscillators characterized by a positive $n$ by $n$ matrix
$\Omega^2$. A local observable of
such a system is a function of the positions and momenta of a
fixed finite set $B$ of oscillators. In this case, $\K=\R^n$,
which I view as $\Ltr(K)$, where $K$ is simply the set of $n$
elements. Indeed, $q\in\R^n$ can be seen as a function
$q:j\in\{1,\dots n\}\mapsto q(j)\in\R$, obviously square
integrable for the counting measure.

Consider now a subset $B$  of $K$, say $B=\{3, 6, 9\}$. A local
observable over $B$ is then a finite linear combination of
operators on $L^2(\R^n)$ of the form $(a_j, b_j\in\R, j\in B)$:
$$
\exp{-\I \left(\sum_{j\in B} (a_jP_j-b_jQ_j)\right)}.
$$
Note that those form an algebra. More generally it is an operator
of the form
$$
\int \left(\Pi_{j\in B} \d a_j \d b_j\right)\ f(a_j, b_j)\
\e^{-\I\left(\sum_{j\in B} (a_jP_j-b_jQ_j)\right)},
$$
for some function $f$ in a reasonable class. In other words, it is
a function of the position and momentum operators of the
oscillators inside the set $B$.

Better yet, if you write, in the Schr\"odinger representation,
$$
L^2(\R^n)\cong L^2(\R^{\sharp B}, \prod_{j\in B}\d x_j)\otimes
L^2(\R^{n-\sharp B}\prod_{j\not\in B}\d x_j),
$$
then it is clear that the weak closure of the above algebra of
local observables is
$$
{\cal B}(L^2(\R^{\sharp B}, \prod_{j\in B}\d x_j))\otimes \bbbone.
$$
So, indeed, a local observable is clearly one that acts only on
the degrees of freedom indexed by elements of $B$. The definition
of a local observable over a finite subset $B$ of any oscillator
lattice is perfectly analogous, where this time the $Q_j, P_j$ are
defined on Fock space, as explained in the previous section. This
definition is natural and poses no problems.

Now what is a local state of such a system? More precisely, I want
to define what a ``strictly local excitation of the vacuum'' is.
The equivalent classical notion is readily described.  The vacuum, being
the ground state of the system, is the quantum mechanical
equivalent of the global equilibrium $X=0$, which belongs of
course to the phase space $\H$, and a local perturbation of this
equilibrium is an initial condition $X=(q,p)$ with the support of
$q$ and of $p$ contained in a subset $B$ of $K$. An example of a
local perturbation of an oscillator lattice is a state $X\in\H$
where only $q_0$ and $p_0$ differ from $0$. In the classical
theory, local perturbations of the equilibrium are therefore
states that differ from the equilibrium state only inside a
bounded subset $B$ of $K$. It is this last formulation that is
readily adapted to the quantum context, through the use of the
notion of ``local observable'' introduced previously. Turning
again to the above example, let me write $|0,\Omega\rangle$ for
the ground state of the oscillator system in the Schr\"odinger
representation; then a strictly local excitation of the vacuum
over $B$ is a state $\psi\in L^2(\R^n, \d x)$ so that, for all
$a_j, b_j\in\R$,
$$
\langle\psi|\exp-\I \sum_{j\not\in  B} (a_jP_j-b_jQ_j)|\psi\rangle
=\langle 0,\Omega| \exp-\I \sum_{j\not\in
B}(a_jP_j-b_jQ_j)|0,\Omega\rangle.
$$
In other words, outside $B$, the states $\psi$ and
$|0,\Omega\rangle$ coincide. Any measurement performed on a degree
of freedom outside $B$ gives the same result, whether the system
is in the vacuum state or in the state $\psi$. The mean kinetic or
potential energy of any degree of freedom outside $B$ is identical
in both cases as well. All this certainly expresses the intuitive
notion of ``localized excitation of the vacuum''. Note that it is
based on the idea of viewing the full system as composed of two
subsystems: the degrees of freedom inside $B$ and the degrees of
freedom outside $B$. The analogous definition of strictly local excitation of the vacuum
for the general class of bose fields considered in the previous section is 
easily guessed and given in Section \ref{s:thingslocal}.

The following question arises naturally. Let's consider a  free bose field and suppose we consider some
state containing one quantum, meaning a state of the form
$a^\dagger(\xi)|0\rangle$. Can such a state be perfectly localized
in a bounded set $B$? Since we like to think of these quanta as
particles, one could a priori expect the answer to be positive,
but the answer is simply: ``NO, not in any model of interest.''
This is the content of the generalization of Knight's theorem proven in \cite{db06} (see Theorem \ref{thm:sumup} (iv)). 
States containing
only one quantum (or even a finite number of them), cannot be
strictly localized excitations of the vacuum. This is a little
surprising at first, but perfectly natural. In fact, it is true
even in finite chains of oscillators, as I will now show.

A special case of the result is indeed easily proven by hand, and
clearly brings out the essential ingredient of the general phenomenon.
Consider a finite oscillator chain, and suppose simply $\Omega^2$
does not have any of the canonical basis vectors $e_i$ of $\R^n$
as an eigenmode. This means that each degree of freedom is coupled
to at least one other one and is certainly true for the
translationally invariant finite chain, to give a concrete
example. I will show by a direct computation that in this
situation there does not exist a one quantum state
$a^\dagger(\xi)|0\rangle$ $(\xi\in\C^n, \overline \xi\cdot
\xi=1)$, that is a perturbation of the vacuum strictly localized
on one of the degrees of freedom, say the first one $i=1$. In
other words, there is no such state having the property that, for
all $Y=(a,b)\in\R^{2n}, a_1=0=b_1$, one has
\begin{equation}\label{eq:locstatfinitedim}
\langle0|a(\xi)\exp-\I \sum_{j=2}^n
(a_jP_j-b_jQ_j)a^\dagger(\xi)|0\rangle =\langle 0|\exp-\I
\sum_{j=2}^n(a_jP_j-b_jQ_j)|0\rangle.
\end{equation}
Now, a simple computation with the Weyl operators shows that this
last condition is equivalent to
$$
\overline\xi\cdot z_\Omega(Y)=0,
$$
for all such $Y$. Here
$z_\Omega(Y)=\frac{1}{\sqrt2}(\Omega^{1/2} a + \I
\Omega^{-1/2}b)$, and simple linear algebra then implies that this
last condition can be satisfied for some choice of $\xi$ if and
only if $e_1$ is an eigenvector of $\Omega$. But this implies it
is an eigenvector of $\Omega^2$, which is a situation I excluded.
Hence $a^\dagger(\xi)|0\rangle$ is not a strictly localized
excitation of the vacuum at site $i=1$ for any choice of $\xi$.

Of course, if the matrix $\Omega^2$ is diagonal, this means that
the degrees of freedom at the different sites are not coupled, and
then the result breaks down. But in all models of interest, the
degrees of freedom at different points in space are of course
coupled. 

The main ingredient for the proof of the generalization of Knight's theorem to free bose fields given in \cite{db06} is the non-locality of $\Omega$. A precise definition will follow below, but the idea is that the operator $\Omega$, which is the
square root of  a finite difference or of a second order
differential operator in all models of interest, does not preserve
supports.  The upshot is
that states of free bose fields with a finite number of particles, and a fortiori,
one-particle states, are never strictly localized in a bounded set
$B$. This gives a precise sense in which the elementary
excitations of the vacuum in a bosonic field theory (relativistic
or not) differ from the ordinary point particles of
non-relativistic mechanics: their Hilbert space of states contains
no states in which they are perfectly localized.

Having decided that one-quantum states cannot be strictly
localized excitations of the vacuum on bounded sets, the question
arises if such strictly localized states exist. Sticking to the simple example of the
chain, any excitation of the vacuum strictly localized on the
single site $i=1$ can be proven to be of the type $\exp \I F(Q_1,
P_1)|0\rangle$, where $F(Q_1, P_1)$ is a self-adjoint operator,
function of $Q_1$ and $P_1$ alone. For a precise statement, see
Theorem \ref{thm:sumup}. Coherent states $\exp -\I
(a_1P_1-b_1Q_1)|0\rangle $ are of this type. So there are plenty such
states. Note however that the linear superposition of two such
states is not usually again such a state: the strictly localized
excitations of the vacuum on the site $i=1$ do not constitute a
vector subspace of the space of all states.  This is in sharp
contrast to what happens when, in the non-relativistic quantum
mechanics of a system of $n$ particles moving in $\R$, we ask the
question: ``What are the states $\psi(x_1, \dots, x_n)$ for which
all particles are in some interval $I\subset \R$?'' These are all
wave functions supported on $I\times\dots\times I$, and they
clearly form a vector space. In that case, to the question ``Are
all particles inside $I$?'' corresponds therefore a projection
operator $P_I$ with the property that the answer is ``yes'' with
probability $\langle\psi|P_I|\psi\rangle$. But in oscillator
lattices there is no projection operator corresponding to the
question ``Is the state a strictly local excitation of the vacuum
inside $B$?''.  This situation reproduces itself in relativistic
quantum field theory, and does not any more constitute a
conceptual problem there as in the finite oscillator chain. I will
discuss this point in more detail in Section \ref{s:reesch}.

In view of the above, it is clear that no position operator for
the quanta of, for example, the Klein-Gordon field can exist.  Those
quanta simply do not have all attributes of the point particles of
our classical mechanics or non-relativistic quantum mechanics
courses. But, since the same conclusion holds for the quanta of a lattice vibration field, this
has nothing to do with causality or relativity, as seems to be generally believed. 
It nevertheless seems that this simple lesson of quantum field theory has met and
still continues to meet with a lot of resistance, as we will see in Section \ref{s:newwigc}. 	

So, to sum it all up, one could put it this way. To the question
\begin{quote}
{Why is there no sharp position observable for  particles?}
\end{quote}
the answer is
\begin{quote}{It is the non-locality of $\Omega$, stupid!}
\end{quote}

More details on this aspect of the story
have been published in \cite{db06}, where it is in particular argued more forcefully that the Newton-Wigner position
operator is not a good tool for describing sharp localization properties of the quanta of quantum fields.



\section{Various viewpoints on localization}\label{s:newwigc}
Now, what is currently the standard view in the physics community
on the question of localization of particles or quanta in field theory? Let me first point out that this is not
necessarily easy to find out from reading the textbooks on quantum
field theory or relativistic quantum physics. Indeed, the least
one can say is that the whole localization issue does not feature
prominently in these books. One seems to be able to detect three
general attitudes. First of all, some books make no mention of it
at all: \cite{ms} \cite{bosh} \cite{ch} \cite{hu} \cite{dereu}. In
\cite{bosh}, for example, when discussing the attributes of particles in an
introductory chapter, the authors identify those as rest mass,
spin, charge and lifetime, but do not mention any notion of
position or localizability.

A second group of authors give an intuitive discussion, based on
the uncertainty principle, together with the non-existence of
superluminal speeds, to explain single massive particles should
not be localizable within a region smaller than their Compton
wavelength: \cite{bd} \cite{sa}  \cite{belipi} are examples. The
idea is that, since $\Delta X\Delta P\geq \hbar$, whenever $\Delta
X$ is of the order of the Compton wavelength $\hbar/mc$, $\Delta
P\geq mc$. Since the gap between positive and negative eigenstates
is $2mc^2$, this is indicating that to obtain such sharp
localization, negative energy eigenstates are ``needed.''   Let me
point out  that this reasoning does not by any means exclude the
possibility of having one-particle states strictly localized in
regions bigger than the Compton wavelength. The argument is then
further used to support the idea that in a fully consistent
relativistic quantum theory, one is unavoidably led to a many body
theory with particle/anti-particle creation, and to field theory.

The essential idea underlying this type of discussion is that,
even though the solutions to the relevant wave equation
(Klein-Gordon or Dirac, mostly) do not, as such, have a
satisfactory probabilistic interpretation, the coupling to other
(classical or quantum) fields is done via the solution and so the
particle is present essentially  where this wave function is not
zero: the particle is where its energy density is. In this spirit,
Bj\"orken and Drell, for example write, at the end of the section
where they address some of the problems associated with the single
particle interpretation of the Klein-Gordon and Dirac equations
(including a brief discussion of the Klein paradox at a potential
step): ``We shall tackle and resolve these questions in Chapter 5.
Before doing this let us look in the vast, if limited, domain of
physical problems where the application forces are weak and
smoothly varying on a scale whose energy unit is $mc^2$ and whose
distance unit is $\frac{\hbar}{mc}$. Hence we may expect to find
fertile fields for application of the Dirac equation for positive
energy solutions.''

This attitude means that whenever a conceptual problem arises, it
is blamed on the single-particle approach taken. The trouble is
that, once these authors treat the full field theory, they do not
come back to the localization issue at all, be it for particle
states or for general states of the field.

A third group of authors give a sligthly more detailed discussion,
including of the Newton-Wigner position operator, but fail to
indicate the problems associated with the latter, leaving the
impression that strict particle localization is quite possible
after all: \cite{sch} \cite{gr} \cite{ste} \cite{stra}. Sterman,
for example, when discussing one-quantum states of the field,
writes: ``To merit the term `particle', however, such excitations
[of the quantum field] should be localisable.'' He then discusses
the Newton-Wigner operator as the solution to this last problem,
without pointing out the difficulties associated with it.
Something similar happens in Schweber who writes: ``By a single
particle state we mean an entity of mass $m$ and spin $0$ which
has the property that the events caused by it are localized in
space.'' Interestingly, neither of them makes any mention of the
contradicting intuitive argument which completely rules out
perfect single particle localization. This is in contrast to
Greiner, who does give this argument,  but does not point the
apparent contradiction with the notion of a position operator, the
(generalized) eigenstates of which are supposed to correspond to a
perfectly localized particle. Of course, this last point is
obscured, as in many cases, by the observation that the
Klein-Gordon wave function corresponding to such a state is
exponentially decreasing with a localization length which is the
Compton wavelength.

Let me note in passing that it is generally admitted that a photon
is not localisable at all, not even approximately. This is
sometimes argued by pointing out it admits no Newton-Wigner
position operator, or intuitively, based on the uncertainty
principle argument which blames this on its masslessness: its
Compton wavelength is infinite. Note however that massless
spinless particles {\em are} localisable in the Newton-Wigner
sense, so that this argument is not convincing. In fact, it was
proven in \cite{bibi} that one-photon states can be localized with
sub-exponential tails in the sense that the field energy density
of the state decreases sub-exponentially away from a localization
center.

To summarize, it seems there is no simple, generally agreed upon
and clearly argued textbook viewpoint on the question of
localization in quantum field theory, even for free fields. The
general idea seems to be that the above intuitive limitations on
particle localization give a sufficient understanding since, at
any rate, the whole issue is not very important. As Bacry puts it,
not without irony, in \cite{ba}: ``The position operator is only
for students and \dots for people interested in the sex of the
angles, this kind of people you find among mathematical
physicists, even among the brightest ones such as Schr\"odinger
and Wigner.''

In my view, it is the absence of a clear definition of ``localized
state'' -- such as the one provided by Knight -- that has left the
field open for competing speculations on how to circumvent the
various problems with the idea of sharp localization of particles
and in particular with the Newton-Wigner operator. Some authors
seem to believe strongly in the need for a position observable for
particles, claiming the superluminal speeds it entails do not
constitute a problem after all, essentially because the resulting
causality violation is too small to be presently observable
\cite{ru3} \cite{flbu}.  Others have provided alternative
constructions with non-commuting components \cite{ba} \cite{flbu}.

To me, this is similar to clinging at all cost to ether theory in
the face of the ``strange'' properties of time implied by
Einstein's special relativity. Between giving up causality or
giving up position operators for field quanta, I have made my
choice. This, together with the definition of Knight and the
accompanying theorem, which could easily be explained in simple
terms in physics books, as Section \ref{s:whethaqua} shows, would
go a long way in clarifying the situation. On top of
that, it would restore the ``democracy between particles''
\cite{ba}. In fact, contrary to what is usually claimed and
although I have not worked out this in detail here, photons are no
worse or better than electrons when it comes to localization, a
point of view that I am not the first one to defend. Peierls, for
example, in \cite{pei}, compares photon and electron properties
with respect to localization and although he starts of with the
statement ``On the other hand, one of the essentially
particle-like properties of the electron is that its position is
an observable, there is no such thing as the position of the
photon,'' he concludes the discussion as follows, after a more
careful analysis of the relativistic regime: ``If we work at
relativistic energies, the electron shows the same disease. So in
this region, the electron is as bad a particle as the photon.''

At any rate, if you find my point of view  difficult to accept and
are reluctant to do so, you are in good company. Here is what
Wigner himself says about it in \cite{wig}, almost forty years
after his paper with Newton: ``One either has to accept this
[referring to non-causality] or deny the possibility of measuring
position precisely or even giving significance to this concept: a
very difficult choice.'' In spite of this, in the conclusion of
this same article, he writes, apparently joining my camp:
``Finally, we had to recognize, every attempt to provide a precise
definition of a position coordinate stands in direct contradiction
to relativity.''

Having advocated Knight's definition of ``strictly local
excitations of the vacuum'', I turn in Section \ref{s:chalocsta}
to their further study. First, we need some slightly more technical material.


\section{All things local}\label{s:thingslocal}
Let's recall we consider harmonic
systems over a real Hilbert space $\K$ of the form $\K = \Ltr(K,
\d \mu)$, where $K$ is a topological space and $\mu$ a Borel
measure on $K$. I
need to give a precise meaning to ``local observables in $B\subset K$'', for every Borel subset of $K$. To do
that, I introduce the notion of ``local structure'', which is a
little abstract, but the examples given below should give you a
good feel for it.
\begin{definition} \label{def:local} A local
structure for the oscillator system determined by $\Omega$ and $\K
= \Ltr(K, \d \mu)$ is a subspace $\S$ of $\K$ with the following
properties:
\begin{enumerate}
\item $\S\subset \K_{1/2}\cap\K_{-1/2}$; \\
\item Let $B$ be a Borel subset of $K$, then $\S_B:=\S\cap \Ltr(B,
\d \mu)$ is dense in $\Ltr(B, \d \mu)$.
\end{enumerate}
\end{definition}
In addition, we need
$$
\H(B,\Omega)\stackrel{\mathrm{def}}{=}\S_B\times \S_B.
$$
Note that, thanks to the density condition in the definition, this
is a symplectic subspace of $\H$. This is a pretty strange
definition, and I will turn to the promised examples in a second,
but let me first show how to use this definition to define what is
meant by ``local observables''. 

\begin{definition} \label{def:localobs}
Let $\K= \Ltr(K, \d \mu), \Omega, \S$ be  as above and let $B$ be
a Borel subset of $K$. The algebra of local observables over $B$
is the algebra
$$
\mathrm{CCR}_0(\H(B,\Omega))=\mathrm{span}\ \{\Wf(z_\Omega(Y))\ | \
Y\in\S_B\times\S_B\}.
$$
\end{definition}
Here ``CCR'' stands for Canonical Commutation Relations. The algebras $\mathrm{CCR}_0(\H(B,\Omega))$ form a net of local algebras
in the usual way \cite{em}  \cite{ho} \cite{ha2}. Note that $\Omega$ plays
a role in the definition of $\S$ through the appearance of the
spaces $\K_{\pm 1/2}$. The first condition on $\S$ guarantees that
$\S\times\S\subset \H$ so that, in particular, for all
$Y\in\S\times\S$, $s(Y,\cdot)$ is well defined as a function on
$\H$ which is  important for the definition of the local
observables to make sense.

For the wave or Klein-Gordon equation, one can choose $\S$ to be either the space of Schwartz functions or $C^\infty_0(\R^d)$. Similarly, on a lattice, one can use the space of sequences of rapid decrease or of finite support. In the simple example of a finite system of oscillators, $\S=\R^n$ will do.

Finally, I need the following definition:
\begin{definition}\label{def:locop}
$\Omega$ is said to be strongly non-local on $B$ if there does not exist a
non-vanishing $h\in\K_{1/2}$ with the property that both $h$ and $\Omega h$
vanish outside $B$.
\end{definition}
Here I used the further definition:
\begin{definition}\label{def:supph} Let $h\in\K_{\pm1/2}$ and $B\subset K$. Then $h$ is said to vanish in $B$ if for all $\eta\in\S_{B}$, $\eta\cdot h=0$. Similarly, it is said to vanish outside $B$, if for all
$\eta\in\S_{B^{\mathrm c}}$, $\eta\cdot h=0$.
\end{definition}
Intuitively, a strongly non-local operator is one
that does not leave the support of any function $h$ invariant. In the examples cited, this is always the case (see \cite{db06bis} for details).

Finally, we can give the general definition of ``strictly local state,'' which goes back to Knight \cite{kn} for relativistic fields, and generalizes (\ref{eq:locstatfinitedim}).
\begin{definition} \label{def:locexc} If $B$ is a Borel subset of $K$, a strictly
local excitation of the vacuum with support in $B$ is a normalized
vector $\psi\in\eff^+(\Kc)$, different from the vacuum itself such that
\begin{equation}\label{eq:localexcitation}
\langle \psi | \Wf(z_\Omega(Y))|\psi\rangle =  \langle 0 |
\Wf(z_\Omega(Y))|0\rangle
\end{equation}
for all $Y=(q,p)\in\H({B^c,\Omega)}$.
\end{definition}
So it is a state which is indistinguishable from the vacuum outside $B$.

\section{Characterizing the stricly local excitations}\label{s:chalocsta}
Having established in Section \ref{s:whethaqua} that in no models of interest finite particle states can be
strictly localized excitations of the vacuum, it is  natural to wonder which
states do have this property. I  mentioned that coherent states are in this
class, as Knight already pointed out. Knight also conjectured that all states
that are strictly localized excitations of the vacuum over some open set $B$,
are obtained by applying a unitary element of the local algebra to the vacuum.
This was subsequently proven for relativistic fields by Licht in \cite{li1}.
Here is a version of this result adapted to our situation \cite{db06bis}.
\begin{theorem} \label{thm:chalocsta}
 Suppose we are given a harmonic system determined by $\Omega$ and
$\K=\L2r(K, \d \mu)$, and with a local structure $\S$.
 Let $B\subset K$ and suppose $\Omega$ is strongly non-local over $B$.
 Let $\psi\in\eff^+(\K^\C)$. Then the following are equivalent:

 (i) $\psi\in\eff^+(\K^\C)$ is a strictly local excitation of the vacuum inside $B$;

 (ii)  There exists a partial isometry $U$, belonging  to the commutant of
 $\mathrm{CCR}_0(\H(B^{\mathrm{c}},\Omega))$ so that
$$
\psi =U|0\rangle.
$$
\end{theorem}
As we have seen, $\Omega$ tends to be strongly non-local over bounded sets in
all examples of interest, so the result gives a complete characterization of
the localized excitations of the vacuum over bounded sets in those cases.

Since the condition in the definition of localized excitation is
qua\-dratic in the state, there is no reason to expect the set of
localized excitations inside $B$ to be closed under superposition
of states. Of course it is closed under the taking of convex
combinations (mixtures). Licht gives a simple criterium in the
cited 1963 paper  allowing to decide whether the linear
combination of two stricly local excitations is still a strictly
local excitation. Both the statement and the proof are again
easily adapted to our situation \cite{db06bis}.

\begin{theorem}\label{thm:conlocsta}  Suppose we are given a
harmonic system determined by $\Omega$ and $\K=\Ltr(K, \d \mu)$,
and with a local
 structure $\S$. Let $B\subset K$ and suppose $\Omega$ is strongly non-local over $B$.
  Let $\psi_1, \psi_2\in\eff^+(\K^\C)$ be strictly local excitations of the vacuum
  inside $B$, so that $\psi_i =U_i|0\rangle$, with
  $U_1, U_2\in\left(\mathrm{CCR}_0(\H(B^{\mathrm{c}},\Omega))\right)'$. Then $$
\psi=(\alpha\psi_1 +\beta\psi_2)/(\parallel
\alpha\psi_1+\alpha\psi_2\parallel)
$$
is a stricly local excitation of the vacuum inside $B$ for all
choices of $(\alpha,\beta)\not=(0,0)$ iff $U_2^{*}U_1$ is a
multiple of the identity operator.
\end{theorem}

The conclusion is then clear. Whenever $\Omega$ is strongly
non-local over $B$, the superposition of strictly localized states
does typically {\em not} yield a strictly localized state over
$B$. In fact, taking $U_1=W_F(z_\Omega(Y_1))$, and
$U_2=W_F(z_\Omega(Y_2))$, with $Y_1, Y_2\in\H(B,\Omega)$, it is
clear that $U_2^*U_1$ is a multiple of the identity only if
$Y_1=Y_2$ so that the localized states do certainly not form a
vector space in that situation. This is in sharp contrast to what
we are used to in the non-relativistic quantum mechanics of
systems of a finite number $N$ of particles. In that case, a wave
function $\psi(x_1, \dots x_N)$ describes a state of the system
with all the particles in a subset $B$ of $\R^3$ iff it vanishes
as soon as one of the variables is outside of $B$. The
corresponding states make up the subspace $L^2(B^N)$ of
$L^2(\R^N)$. In that case, to the question ``Are all the particles
in the set $B$?'' corresponds a projection operator $P_B$ with the
property that the answer is ``yes''  with probability $\langle
\psi|P_B|\psi\rangle$. To the question ``Is the state a stricly
local excitation of the vacuum in $B$?'' cannot correspond such a
projection operator! I will belabour this point in Section
\ref{s:reesch}.

\section{Surprises?}\label{s:reesch}
Here is  a list of three mathematical truths that have originally been proven
in the context of relativistic quantum field theory, and that seem to have
generated a fair amount of surprise and/or debate:
\begin{enumerate}
\item The vacuum is a cyclic vector for the local algebras over open  regions.

\item The vacuum is  a separating vector for the local algebras over open  regions.

\item The set of stricly local states (in the sense of Knight)
over an open region is not closed under superposition of states.
\end{enumerate}
In that context, the open regions referred to are open regions of
Minkowski space time. And by relativistic, I mean of course
invariant under the Poincar\'e group. Recall that a vector $\phi$ in a Hilbert space $\mathcal V$ is cyclic for an algebra $\mathcal A$ of bounded operators on $\mathcal V$ if $\overline{\mathrm{span}\mathcal A \phi}=\mathcal V$. And it is separating if $A\in\mathcal A$, $A\phi=0$, implies $A=0$. 

It turns out that, as soon as $\Omega$ is a
non-local operator, suitably adapted analogous statements hold for
the harmonic systems that are the subject of this paper. We already saw this for the third statement in the previous section. The
results are summed up in the following theorem (proven in \cite{db06bis}).
\begin{theorem}\label{thm:sumup}
Let $\K=L^2(K, \d \mu)$, $\Omega^2\geq0$ and $\S$ be as before.
Suppose that, for some $B\subset K$, $\Omega$ is strongly
non-local over $B$. Then

(i) The vacuum is a cyclic vector for
$\mathrm{CCR}_0(\H(B^{\mathrm{c}},\Omega))$;

(ii) The vacuum is a separating vector for
$\mathrm{CCR}_\mathrm{w}(\H(B,\Omega))$;

(iii) The set of stricly local states over $B$ do NOT form a
vector space.

(iv) There do not exist finite particle states that are stricly
local states over $B$.

\end{theorem}
Since, as we pointed out, the hypotheses of the theorem hold in
large classes of examples, and for various sets $B$, so do the
conclusions. Note that they therefore do not have a particular
link with relativistic invariance. Note also that for lattices,
the vacuum cannot be cyclic for a bounded set, since then
$\H(B,\Omega)$ is finite dimensional so that
$\overline{\mathrm{span}_\C} z_\Omega(\H(B,\Omega))$ is a strict
subspace of $\K^\C$. However, it is easily shown in translationally
invariant models, for example, that the vacuum is cyclic for the the CCR-algebra
over the complement of any bounded set $B$ (See \cite{db06bis} for details).

My goal in this section is to  explain in each case why the above statements
have generated surprise in the context of relativistic field theory, then to
argue that none of these properties should have surprised anyone precisely
since they hold for simple systems of $n$ coupled oscillators, and for free
bose fields in rather great generality as the previous theorem shows. In
particular, they therefore hold for the Klein-Gordon equation on Minkowski
spacetime, which happens to be relativistic (meaning here Poincar\'e invariant). Why should it then be
contrary to anyone's physical intuition if they continue to hold for
interacting relativistic fields?

{\bf Statement 1.}  This was proven for relativistic quantum
fields by Reeh and Schlieder in \cite{resc} and has been  a
well-known feature of {\em relativistic} quantum field theory ever
since. As already mentioned, in that context, the open regions
referred to are open regions of Minkowski space time. For a
textbook formulation in the context of axiomatic, respectively
algebraic relativistic quantum field theory you may consult
\cite{stwi}, respectively \cite{ha2} \cite{ho}. 

Rather than giving the full proof of Theorem \ref{thm:sumup} (i),  I will  once again restrict myself to a system of $n$ oscillators and consider the subspace of
the state space $L^2(\R^n)$ containing all vectors of the form
$$
\sum_{j=1}^L c_j\exp -\I (a_jP_1-b_jQ_1)|0,\Omega\rangle,
$$
for all choices of $c_1\dots c_L,\ L\in\N$. Note that, since only the operators $Q_1$ and $P_1$ occur in the exponents, it follows from the considerations of the previous sections that such vector is a
linear combination of excitations of the vacuum that are strictly
localized at site $i=1$. Nevertheless, one can easily prove that
that those vectors form a dense subset of the full Hilbert space, which means that the vacuum is a cyclic vector for the local algebra over that site.
A very pedestrian proof goes as follows: thinking for simplicity
of the case $n=2$, it is easy to see, taking limits, that the
vectors $Q_1^k|0, \Omega\rangle$ and $P_1^\ell|0,\Omega\rangle$
belong to the closure of the span of the above vectors. Now, using
that $\Omega$ is not diagonal, one easily concludes that therefore
all wave functions of the form
$$
p(x_1, x_2)\exp-\frac12 x\cdot \Omega x,
$$
with $p(x_1,x_2)$ any polynomial belong to the space. Taking
Hermite polynomials, for example, one obtains a basis for the full
state space $L^2(\R^2)$. Note that, again, $\Omega$ has to be
non-diagonal for this to work. So indeed, the vacuum is a cyclic vector for the algebra of local
observables over $B$ (here $B=\{1\}$). It holds in much more
generality, provided $\Omega$ is non-local: this is the content of
Theorem
\ref{thm:sumup}. There are various poetic and misleading ways to
express this result, for example by saying: ``Local operations on
the vacuum can produce instantaneous and arbitrary changes to the
state vector arbitrarily far away.'' Lest one enjoys confusing oneself, it is a good idea to stay
clear of such loose talk, as I will further argue below.

Let me now
corroborate my claim that the cyclicity of the vacuum came as a surprise when it
was proven for relativistic fields. Segal writes in \cite{se4} it is ``particularly
striking'' and Segal and Goodman say it is ``quite surprising''
and a ``bizarre phenomenon'' in \cite{sego}. Streater and Wightman
call it ``a surprise'' in \cite{stwi}. Even rather recently, Haag
refers to it as ``startling'' in \cite{ha2} (p. 102), although he
reduces this qualification to ``(superficially) paradoxical''
later on in his book (p. 254).  Redhead in \cite{re} similarly
calls it ``surprising, even paradoxical''.

The surprise finds its origin in an apparent contradiction between
the above mathematical statement and some basic physical intuition
on the behaviour of quantum mechanical systems. Segal for example
writes in \cite{se4} that Reeh-Schlieder is particularly striking
because it apparently means that ``the entire state vector space
of the field could be obtained from measurements in an arbitrarily
small region of space-time.'' This, he argues, is ``quite at
variance with the spirit of relativistic causality.'' Similar
arguments can be found for example in \cite{re} or in \cite{ha2}
and in \cite{flbu}, the authors write that ``it is hard to square
with na\"\i ve, or even educated, intuitions about localization.''

This supposed contradiction is, as we shall see, directly related
to the misconception of the vacuum as ``empty space,'' which
already was part of the problem with the debate surrounding the
Newton-Wigner position operator. To understand this, let me
explain what the contradiction translates to in  our present
context of harmonic systems: it is the too naive and, as we shall
see, erroneous expectation that the mathematical operation of
applying to the vacuum vector a local observable $A$ belonging to
the local algebra over some subset $B$ of $K$  yields a state
$A|0\rangle/\langle0|A^* A|0\rangle^{1/2}$ which is a strictly
localized excitation of the vacuum in $B$  in the sense of
Knight's Definition \ref{def:locexc} (for brevity called ``local states''
in what follows) \cite{hasw}. This, if it were true,  would  of
course be in blatant contradiction with the cyclicity of the
vacuum. Indeed, if the vacuum is cyclic, any state of the system,
including one that differs from the vacuum very far away from $B$
can be approximated by one of the above form. Fortunately, we know
from Licht's result that applying a local observable to the vacuum
yields a stricly local excitation only if the local observable is
a partial isometry. Of course, this only shifts the paradox,
because one may choose to find Licht's result paradoxical. Indeed,
when $A$ is a projector, one can, according to the standard
interpretational rules of quantum mechanics, and in particular the
``collapse of the wave function'' prescription, prepare (in
principle!) an ensemble of systems, all in the state
$A|0\rangle/\langle0|A^* A|0\rangle^{1/2}$.  Now, if the projector
$A$ is a local observable, these measurements can correctly be
thought of as being executed within $B$. Now, {\em if you think of
the vacuum as empty space}, it is inconceivable on physical
grounds that such a measurement could instantaneously change
something outside $B$. But this is then in contradiction with the
mathematical result of Licht which asserts that if $A$ is a
projector, the state $A|0\rangle/\langle0|A^* A|0\rangle^{1/2}$ is
{\em not} local and therefore {\em does} differ from the vacuum
outside $B$.  The way out is obvious: the vacuum is not empty
space but the ground state of an extended system. To see what is
happening, the  example of the chain of $n$ coupled oscillators of
is again instructive. Concentrate for
example on the seventh oscillator of the chain and consider the
question: ``Does the displacement of the seventh oscillator fall
within the interval $[a,b]$?'' To this corresponds the projector
$\chi_{[a,b]}(Q_7)$. The outcome of the corresponding preparation
procedure will be an ensemble of systems, all in the  state
$$
\chi_{[a,b]}(Q_7)|0\rangle/\langle0|\chi_{[a,b]}(Q_7)|0\rangle
$$
with the non-vanishing probability $\langle0|\chi_{[a,b]}(Q_7)|0\rangle$. But
it is obvious that this state differs from the vacuum on the neighbouring site
$8$ and even on very far away sites! So even though the above projector
corresponds to a local physical operation it does nevertheless not lead to a
local excitation of the vacuum
  because even a local physical operation (here a measurement of the displacement of
a single oscillator on one site) on the vacuum will
instantaneously change the state of the system everywhere else.
This is obviously the result here of the fact that the vacuum
exhibits correlations between (commuting!) observables at
different sites along the ring, a perfectly natural and expected
phenomenon. After all, the oscillators on different sites are
connected by springs. The ultimate reason for this phenomenon is
therefore that the ground state of a typical oscillator system
characterized by an $n$ by $n$ matrix $\Omega^2$ is an
``entangled'' state in $L^2(\R^n)\cong L^2(\R)\otimes
L^2(\R)\dots\otimes L^2(\R)$, unless of course $\Omega^2$ is
diagonal so that the oscillators are uncoupled to begin with. This
entanglement can be seen in the fact that the ground state is a
Gaussian with correlation matrix $\Omega$, which is not diagonal.
The change far away that the vacuum undergoes in the above
measurement process is therefore nothing new, but a version of the
usual ``weirdness'' of quantum mechanics, at the origin also of
the EPR paradox.

In conclusion, I would therefore like to claim that in oscillator
systems such as the ones under study here, one should
expect that physical changes (such as measurements) operated on
the vacuum vector inside some set $B\subset K$ (for example as the
result of a measurement) will alter the state of the system
outside $B$ instantaneously. This is already true for finite
dimensional systems as explained above and  remains true for
infinite dimensional ones such as oscillator lattices or the
Klein-Gordon equation. That the latter has the additional feature
of being Poincar\'e invariant does not in any way alter this
conclusion nor does it lead to additional paradoxes. It does in particular
not cause any causality problems in the sense that the phenomenon
cannot be used to send signals, as one can easily see.

In short, the cyclicity of the vacuum does not lead to any
contradictions with basic physical intuition, provided the latter
is correctly used. It is perhaps interesting to note that Licht's
1963 result which is of great help in understanding the situation,
is not cited in any of the other works on the subject I mentioned
(although his paper sometimes is to be found in the bibliography
\dots).

{\bf Statement 2.} What about the separability of the vacuum?  The separability  also gives rise to
an apparent contradiction that is equally easily dispelled with. Indeed, the
``surprise'' can be formulated as follows. If a non-trivial projector $P$
belongs to the local algebra CCR$_w(\H(B, \Omega)$, then $P|0\rangle\not=0$
since the vacuum is a separating vector. Now this means that if the system is
in its vacuum state and one measures locally some property of the system, such
as, for example, whether the displacement of the oscillator on site $69$  has
a value between $7$ and $8$, then the answer is ``yes'' with non-zero
probability. This consequence of separability (possibly first mentioned in
\cite{hekr}) can be paraphrased suggestively as follows:
\begin{quotation}``When the system is in the vacuum state,
anything that can happen will happen,''
\end{quotation}
or, alternatively, as in \cite{suwe}, ``every local detector has a
non-zero vacuum rate.'' This result is certainly paradoxical if
you think of the vacuum as being empty space. Indeed, how can any
measurement, local or not, give a non-trivial result in empty
space?

To see why there is no reason to be surprised, let us look again
at my favourite example, a system of $n$ coupled oscillators. Its
ground state is a Gaussian with correlation matrix $\Omega$, so the probability that the displacement of
the oscillator on site $69$  has a value between $7$ and $8$ is
obviously non-zero! That a similar property survives in the
quantized Klein-Gordon field does not strike me as particularly
odd, since the mathematical structure of both models is exactly
identical, as should be clear from the previous sections.

{\bf Statement 3.} In the context of relativistic quantum field theory, this
was proven by Licht in his cited 1963 paper. He uses as a basic ingredient the
result of Reeh and Schlieder. In our context here, it is the content of
Theorem \ref{thm:conlocsta}, which is a consequence of the strong non-locality
of $\Omega$.

Now, Theorem \ref{thm:conlocsta} has an interesting consequence
for quantum measurement theory. Indeed, given a set $B$ so that
$\Omega$ is strongly non-local over $B$, there cannot exist a
projection operator $P_B$ on $\eff^+(\K^\C)$ with the property
that $P_B\psi=\psi$ if and only if $\psi\in\eff^+(\K^\C)$ is a
strictly local excitation of the vacuum over $B$. Indeed, if such
an operator existed, the strictly localized excitations would be
stable under superposition of states, of course. So there is no
projector associated to the ``yes-no'' question: ``Is the system
strictly localized in $B$?''  This is different from what we are
used to in the non-relativistic quantum mechanics of  a finite
number of particles. There questions such as ``Are all particles
in $B$?'' have a projector associated to them. But of course, we
are dealing here with extended systems, such as oscillator
lattices, and asking questions about excitations of the vacuum,
not about the whereabouts of the individual oscillators, for
example! It is only when you forget that, and try to interpret all
statements about the fields in terms of particles that you run
into trouble with your intuition.

As I pointed out before, this third statement above follows from the second,
the second from the first and the main ingredient of the proof of the first by
Reeh and Schlieder is relativistic invariance and the spectral property. This
seems to have lead to the impression that these three properties, and
especially the last one, are typical of relativistic fields, and absent in
non-relativistic ones. For example, Redhead  says in \cite{re} that ``to
understand why the relativistic vacuum behaves in such a remarkable way, let
us begin by contrasting the situation with nonrelativistic quantum field
theory''. It is furthermore said in \cite{hasw} that ``in a relativistic field
theory it is not possible to define a class of states strictly localized in a
finite region of space within a given time interval if we want to keep all the
properties which one would like to associate with localization.'' The authors
include in those properties the fact that it has to be a linear manifold.
Quite recently still, a similar argument is developed in some detail
 in \cite{by}.
They explain that ``there are marked differences between non-relativistic and
relativistic theories, which manifest themselves in the following alternative
structure of the set of vectors'' $\psi$ representing states that are strictly
localized excitations of the vacuum. They then go on to explain that, in the
non-relativistic case, the set of localized states are closed under
superposition,  whereas in relativistic quantum field theory, they are not.

But these statements are potentially misleading since  precisely
the same phenomena produce themselves in the eminently
non-relativistic systems of coupled oscillators that I have been
describing as is clear from the results of the previous sections.
These phenomena are a consequence of the strong non-locality of
$\Omega$, and have nothing to do with relativistic invariance.
 The problem is  that one has the tendency to
compare relativistic field theories with the second quantization
of the Schr\"odinger field, which is of course a non-relativistic
field theory. In that context, the above three statements do not
hold and in particular, the set of local states is a linear
subspace of the Hilbert space. But if you compare, as you should,
relativistic field theories to the equally non-relativistic
harmonic systems, such as lattices of coupled oscillators, you
remark that many of the features of the relativistic fields are
perfectly familiar from the non-relativistic regime. They should
therefore not come as a surprise, and not generate any paradoxes.
 It should be noted  that already in \cite{sego} the
source of the Reeh-Schlieder properties for free relativistic
fields is identified to be the non-locality of $(-\Delta
+m^2)^{1/2}$. This is further exploited in \cite{mas1} \cite{mas2}
and \cite{ve} where the anti-locality of potential perturbations
of $-\Delta$, respectively of the Laplace-Beltrami operator on
Riemannian manifolds is proven, which then yields a proof of the
Reeh-Schlieder property in these situations.

As a further remark along those lines, I would like to point out that the fact
that
 the stricly local states are not stable under superposition of states is
sometimes related to the type of the local algebras of observables
\cite{li1} \cite{by}. In relativistic quantum field theory, they
are known to be of type III \cite{dr}, a result that generated a
fair amount of excitement when it was discovered, since type III
factors were thought to be esoteric objects \cite{se4}. It should
be noted, however, that the local algebras of observables in
oscillator lattices are type I factors, and that the stricly local
excitations nevertheless are not stable under superposition. In
addition, just as in relativistic quantum field theory, pure
states  look locally like mixtures: this is a consequence of
entanglement, not of relativity.

A further paradox related to the third statement is the following.
Suppose you have a strictly local excitation of the vacuum $\psi$
over some set $B$. Now ask yourself the question if a local
measurement inside $B$ can prepare this state. In other words, is
the projector onto $\psi$ a local observable? In relativistic
theories, the answer is in the negative \cite{re}. Indeed, since
the algebras are of type III, they contain no finite dimensional
projectors. But even in oscillator lattices, the answer is
negative, since the local algebras do not contain finite
dimensional projectors either, and this despite the fact that they
are of type I. Of course, this is not in the least little bit
surprising: intuitively also, to fix the state of an extended
system such as an oscillator lattice, you expect to need to  make
measurements on every site of the lattice. In particular, if the
state is a stricly local excitation of the vacuum on the fifth
site, you should check it coincides
 with the vacuum on all other sites. So you can neither measure nor prepare such a
 state by working only on a few lattice sites. Again, the situation is very different
 from the one of, for example, a one-electron system, where local states can be
 prepared locally.

 Of course, by now, I hope I have brainwashed you into agreeing
 that it is really a bad idea to think of the vacuum as empty
 space. But should you not be convinced, again, you are in
 excellent company. This is how Schwinger talks about the vacuum in
 \cite{schwi}:
  ``With [quantum field theory] the vacuum becomes once again a physically
reasonable state with no particles in evidence. The picture of an
infinite sea of negative energy electrons is now but regarded as
an historical curiosity, and forgotten. Unfortunately, this
episode, and discussions of vacuuum fluctuations, seem to have
left people with the impression that the vacuum, the physical
state of nothingness (under controlled physical circumstances), is
actually the scene of wild action.'' And a bit further down in the
same article, he insists again: ``I recall that for us the vacuum
is the state in which no particles exist. It carries no physical
properties: it is structureless and uniform. I emphasize this by
saying that the vacuum is not only the state of minimal energy, it
is the state of {\em zero} energy, zero momentum, zero charge,
zero whatever. Physical properties, structure, come into existence
only when we disturb the vacuum, when we excite it.''


\section{Conclusions}
As long
as one studies only a finite number of oscillators, the imaginative description of the quantum states of harmonic systems  in terms of quanta is rather cute but not terribly useful or important. It is however a crucial
element of relativistic and non-relativistic quantum field theories, which have an infinite
number of degrees of freedom. In that case, the quanta
are traditionally interpreted as particles. Photons, for example, are the quanta of the
electromagnetic field and phonons are those of the vibration field.
Electrons are similarly quanta of the Dirac field.

Now if you want to interpret the quanta as particles, you automatically are
lead to the question that features as the title of this manuscript. One thinks of
a particle as a
localized object, and so it seems perfectly natural to wish to have a position operator for it,
or at least some way to answer questions such as : ``What is the probability
of finding the particle in such and such a region of space?'' In fact, as I
have argued via the generalization of Knight's theorem, since the particles of quantum field theory are quanta, the
situation is similar to the one we discovered already with
finite systems of oscillators: the quanta cannot
be perfectly localized and therefore there is no way to associate a position
operator to them, and in that sense the question above does not really make any sense at all. It should be noted that whereas Knight's definition is regularly referred to in discussions of localization issues, his theorem, which is very helpful in understanding the issues at hand, seems to never be mentioned.

That quanta cannot be perfectly localized does not constitute a problem. 
A good notion of localized states exist: it is the one provided by considering 
localized excitations of the vacuum and goes back to Knight. Those states differ from the vacuum only inside a set $B$ and there are plenty of them. They do however not form a vector subspace of the quantum Hilbert space, and no projection operator is associated with the localized excitations over a fixed set $B$. As I have explained, this feature of relativistic quantum field theories is also familiar from non-relativistic oscillator chains, and as such not related to relativistic invariance. If the right analogies between relativistic and non-relativistic theories are used, it is not counter-intuitive or in any way surprising.

Let me mention that a complete discussion of the localization properties of field states should also analyse notions of approximate localization, allowing for exponential or algebraic decay of the expectation values outside the set $B$. These notions are implicit in the physics literature, where states with exponential tails, for example, are often thought of as localized. They have been developed by several authors \cite{am} \cite{hasw} \cite{bibi} \cite{ha2} \cite{wa1}. A complete discussion should finally  also address those questions for other fields, such as complex bose fields and Fermi fields. These issues will be adressed elsewhere \cite{tis}.

In conclusion, the morale of the story is this: when testing your understanding of a notion
in quantum field theory, try to see what it gives for a finite system of
oscillators. Examples of situations where this algorithm seems to meet with some success are the various puzzles associated with particle localization and the Reeh-Schlieder theorem and its consequences, as I have argued here.
In particular, if the notion under study, when adapted to finite or infinite oscillator chains, looks funny there, it is likely to lead you astray in the
context of quantum field theory as well: an example is the Newton-Wigner position operator. 
Of course, I am not the first one to point these analogies out. In \cite{pei}, one can read:
``The radiation field differs from atomic systems principally by
\dots having an infinite number of  degrees of freedom. This may
 cause some difficulties in visualizing the
physical problem, but is not, in itself, a
difficulty of the formalism.''



\bibliographystyle{alpha}

\bibliography{biblio}

\begin{thebibliography}{Wal01b}

\bibitem[Amr69]{am}
W.~Amrein.
\newblock Localizability for particles of mass zero.
\newblock {\em Helv. Phys. Acta}, 42:149--190, 1969.

\bibitem[Bac88]{ba}
H.~Bacry.
\newblock {\em Localizability and space in quantum physics}.
\newblock Lecture Notes in Physics 308, Springer Verlag. 1988.

\bibitem[BB98]{bibi}
I.~Bialynicki-Birula.
\newblock Exponential localization of photons.
\newblock {\em Phys. Rev. Lett.}, 80(24):5247--5250, 1998.

\bibitem[BD65]{bd}
J.~D. Bj\"orken and S.~D. Drell.
\newblock {\em Relativistic quantum fields}.
\newblock Mc. Graw Hill, NY. 1965.

\bibitem[Bi{\`e}]{db06bis}
S.~De Bi{\`e}vre.
\newblock {\em Classical and quantum harmonic systems (in preparation)}.

\bibitem[Bi{\`e}06]{db06}
S.~De Bi{\`e}vre.
\newblock Local states of free bose fields.
\newblock In {\em Large Coulomb Systems}, volume 695 of {\em Lecture Notes in
  Physics}, pages 17--63. 2006.

\bibitem[BLP71]{belipi}
V.B. Berestetskii, E.M. Lifshitz, and L.P. Pitaevskii.
\newblock {\em Relativistic quantum theory}.
\newblock Pergamon Press, Oxford. 1971.

\bibitem[BS83]{bosh}
N.~N. Bogoliubov and D.V. Shirkov.
\newblock {\em Quantum fields}.
\newblock The Benjamin/Cummings Publishing Company Inc. 1983.

\bibitem[BY94]{by}
D.~Buchholz and J.~Yngvason.
\newblock There are no causality problems for fermi's two-atom system.
\newblock {\em Physical Review Letters 5}, 73:613--616, 1994.

\bibitem[Cha90]{ch}
S.J. Chang.
\newblock {\em Introduction to quantum field theory}.
\newblock World Scientific. 1990.

\bibitem[Der01]{dereu}
J.P. Dereudinger.
\newblock {\em Theorie quantique des champs}.
\newblock Presses Polytechniques et Universitaires Romandes. 2001.

\bibitem[Dri75]{dr}
W.~Driessler.
\newblock Comments on lightlike translations and applications in relativistic
  quantum field theory.
\newblock {\em Commun. Math. Phys.}, 44(2):133--141, 1975.

\bibitem[Emc72]{em}
G.~Emch.
\newblock {\em Algebraic methods in statistical mechanics and quantum field
  theory}.
\newblock Wiley Interscience, 1972.

\bibitem[FB99]{flbu}
G.~Fleming and J.~Butterfield.
\newblock Strange positions.
\newblock In {\em From physics to philosophy}, pages 108--165. Cambridge
  University Press, 1999.

\bibitem[Gre90]{gr}
W.~Greiner.
\newblock {\em Relativistic quantum mechanics}.
\newblock Springer. 1990.

\bibitem[Haa96]{ha2}
R.~Haag.
\newblock {\em Local quantum physics}.
\newblock Springer. 1996.

\bibitem[Hal01]{hal1}
Hans Halvorson.
\newblock Reeh-{S}chlieder defeats {N}ewton-{W}igner: on alternative
  localization schemes in relativistic quantum field theory.
\newblock {\em Philos. Sci.}, 68(1):111--133, 2001.

\bibitem[HC02]{clha2}
Hans Halvorson and Rob Clifton.
\newblock No place for particles in relativistic quantum theories?
\newblock {\em Philos. Sci.}, 69(1):1--28, 2002.

\bibitem[Heg74]{he1}
G.~C. Hegerfeldt.
\newblock Remark on causality and particle localization.
\newblock {\em Phys. Rev. D}, 10(10):3320--3321, 1974.

\bibitem[Heg85]{he2}
G.~C. Hegerfeldt.
\newblock Violation of causality in relativistic quantum theory?
\newblock {\em Phys. Rev. Letters}, 54(22):2395--2398, 1985.

\bibitem[Heg98]{he3}
G.~C. Hegerfeldt.
\newblock Causality, particle localization and positivity of the energy.
\newblock In {\em Irreversibility and causality (Goslar, 1996)}, volume 504 of
  {\em Lecture Notes in Phys.}, pages 238--245. Springer, Berlin, 1998.

\bibitem[HK70]{hekr}
K.-E. Hellwig and K.~Kraus.
\newblock Operations and measurements {I}{I}.
\newblock {\em Commun. Math. Phys.}, 16:142--147, 1970.

\bibitem[Hor90]{ho}
S.~S. Horuzhy.
\newblock {\em Introduction to algebraic quantum field theory}, volume~19 of
  {\em Mathematics and its Applications (Soviet Series)}.
\newblock Kluwer Academic Publishers Group, Dordrecht, 1990.

\bibitem[HS65]{hasw}
R.~Haag and J.~A. Swieca.
\newblock When does a quantum theory describe particles?
\newblock {\em Commun. Math. Phys.}, 1:308--320, 1965.

\bibitem[Hua98]{hu}
K.~Huang.
\newblock {\em Quantum field theory}.
\newblock J. Wiley and Sons, 1998.

\bibitem[Kni61]{kn}
J.M. Knight.
\newblock Strict localization in quantum field theory.
\newblock {\em Journal of Mathematical Physics}, 2(4):459--471, 1961.

\bibitem[Lic63]{li1}
A.~L. Licht.
\newblock Strict localization.
\newblock {\em Journal of Mathematical Physics}, 4(11):1443--1447, 1963.

\bibitem[Mas68]{mas1}
K.~Masuda.
\newblock A unique continuation theorem for solutions of the wave equation with
  variable coefficients.
\newblock {\em J. Math. Analysis and Applications}, 21:369--376, 1968.

\bibitem[Mas73]{mas2}
K.~Masuda.
\newblock Anti-locality of the one-half power of elliptic differential
  operators.
\newblock {\em Publ. RIMS, Kyoto University}, 8:207--210, 1972/73.

\bibitem[MG84]{ms}
F.~Mandl and G.Shaw.
\newblock {\em Quantum field theory}.
\newblock Wiley. 1984.

\bibitem[Pei73]{pei}
R.E. Peierls.
\newblock The development of quantum field theory.
\newblock In {\em The physicist's conception of nature}, pages 370--379. D.
  Reidel Publishing Cy., Dordrecht, Netherlands, 1973.

\bibitem[Red95]{re}
M.~Redhead.
\newblock More ado about nothing.
\newblock {\em Foundations of Physics 1}, 25:123--137, 1995.

\bibitem[RS61]{resc}
H.~Reeh and S.~Schlieder.
\newblock Bemerkungen zur {U}nit\"ar\"aquivalenz von {L}orentzinvarianten
  {F}elden.
\newblock {\em Nuovo cimento (10)}, 22:1051--1068, 1961.

\bibitem[Rui81]{ru3}
S.~Ruijsenaars.
\newblock On newton-wigner localization and superluminal propagation speeds.
\newblock {\em Ann. Physics}, 137(1):33--43, 1981.

\bibitem[Sak67]{sa}
J.J. Sakurai.
\newblock {\em Advanced quantum mechanics}.
\newblock Addison Wesley Publishing Company. 1967.

\bibitem[Sch61]{sch}
S.~S. Schweber.
\newblock {\em An introduction to relativistic quantum field theory}.
\newblock Row and Peterson and Company, 1961.

\bibitem[Sch73]{schwi}
J.~Schwinger.
\newblock A report on quantum electrodynamics.
\newblock In {\em The physicist's conception of nature}, pages 413--429. D.
  Reidel Publishing Cy., Dordrecht, Netherlands, 1973.

\bibitem[Seg63]{se4}
I.~E. Segal.
\newblock Quantum fields and analysis in the solution manifolds of differential
  equations.
\newblock In {\em Proc. Conf. on Analysis in Function Space}, pages 129--153.
  M.I.T. Press, Cambridge, 1963.

\bibitem[SG65]{sego}
I.~E. Segal and R.~W. Goodman.
\newblock Anti-locality of certain {L}orentz-invariant operators.
\newblock {\em J. Math. Mech.}, 14:629--638, 1965.

\bibitem[Ste93]{ste}
G.~Sterman.
\newblock {\em An introduction to quantum field theory}.
\newblock Cambridge University Press, 1993.

\bibitem[Str98]{stra}
P.~Strange.
\newblock {\em Relativistic quantum mechanics}.
\newblock Cambridge University Press. 1998.

\bibitem[SW64]{stwi}
R.~F. Streater and A.~S. Wightman.
\newblock {\em PCT, spin and statistics and all that}.
\newblock The Benjamin/Cummings Publishing Cy, Reading MA, 1964.

\bibitem[SW85]{suwe}
S.~Summers and R.~Werner.
\newblock The vacuum violates {B}ell's inequalities.
\newblock {\em Physics Letters 5}, 110A:257--259, 1985.

\bibitem[Tis]{tis}
N.~Tisserand.
\newblock {\em PhD thesis (in preparation)}.

\bibitem[Ver94]{ve}
R.~Verch.
\newblock Anti-locality and a {R}eeh-{S}chlieder theorem on manifolds.
\newblock {\em Letters in Mathematical Physics}, 28(2):143--154, 1994.

\bibitem[Wal01a]{wa2}
D.~Wallace.
\newblock Emergence of particles from bosonic quantum field theory.
\newblock {\em preprint 2001, http://arxiv.org/abs/quant-ph/0112149}, 2001.

\bibitem[Wal01b]{wa1}
D.~Wallace.
\newblock In defence of naivet{\'e}: The conceptual status of lagrangian
  quantum field theory.
\newblock {\em preprint 2001, http://arxiv.org/abs/quant-ph/0112148}, 2001.

\bibitem[Wig83]{wig}
E.~P. Wigner.
\newblock Interpretation of quantum mechanics.
\newblock In {\em Quantum theory and measurement}, pages 260--314. Princeton
  University Press, Princeton, NJ, 1983.

\end{thebibliography}

\end{document}